\documentclass[reprint, amsmath,amssymb,aps ,prl 
]{revtex4-1}

\usepackage{graphicx}
\usepackage{dcolumn}
\usepackage{bm}
\usepackage{hyperref}

\def\CO{{\cal O}}

\def\CW{{\cal W}}

\def\CT{{\cal T}}

\DeclareMathOperator{\Time}{T}

\newcommand{\vev}[1]{ \left\langle {#1} \right\rangle }
\newcommand{\bra}[1]{ \langle {#1} | }
\newcommand{\ket}[1]{ | {#1} \rangle }

\def\Im{\mathop{\rm Im}}
\def\Re{\mathop{\rm Re}}

\def\beq#1\eeq{\begin{align}#1\end{align}}

\begin{document}


\title{Unitarity, Locality, and Scale versus Conformal Invariance in Four Dimensions}

\author{Kazuya Yonekura}
\affiliation{School of Natural Sciences, Institute for Advanced Study, 1 Einstein Drive, Princeton, NJ 08540, USA.}

\begin{abstract}
In four dimensional unitary scale invariant theories, arguments based on the proof of the $a$-theorem suggest that the trace of the 
energy-momentum tensor $T$ vanishes when the momentum is light-like, $p^2=0$.
We show that there exists a local operator $O$ such that the trace is given as $T=\partial^2 O$, which establishes the equivalence
of scale and conformal invariance. We {\it define} the operator as $O=\partial^{-2} T$, and explain why this is a well-defined local operator.
Our argument is based on the assumptions that: (1) A kind of crossing symmetry for vanishing matrix elements holds regardless of the existence
of the S-matrix. (2) Correlation functions in momentum space are analytic functions other than singularities and branch cuts coming from on-shell processes.
(3) The Wightman axioms are sufficient criteria of the locality of an operator.
\end{abstract}

\maketitle


\paragraph{Introduction.}---
One of the long standing problems in quantum field theory is whether unitary scale invariant theory is conformally invariant or not.
This is a highly nontrivial question since there is a simple non-unitary counterexample \cite{Riva:2005gd}.
The equivalence of scale and conformal invariance in unitary theory is proved long time ago in two dimensions~\cite{Zamolodchikov:1986gt,Polchinski:1987dy}.
Evidence has been accumulated
\cite{Jack:1990eb,Jack:1990eb,Osborn:1991gm,Dorigoni:2009ra,Nakayama:2010wx,ElShowk:2011gz,Antoniadis:2011gn,Zheng:2011bp,Luty:2012ww,
Fortin:2012hn,Fortin:2012hc,Nakayama:2012nd,Yonekura:2012kb,Dymarsky:2013pqa,Farnsworth:2013osa,Baume:2014rla,
Bzowski:2014qja,Dymarsky:2014zja,Sachs:2014cga}
which support the equivalence
in other dimensions at least under certain conditions. See \cite{Nakayama:2013is} for a review.

In particular, in four dimensions, there is a very strong nonperturbative argument \cite{Luty:2012ww} 
(see also \cite{Nakayama:2011wq}) 
which follows the proof of  the $a$-theorem
using dilaton scattering amplitudes \cite{Komargodski:2011vj,Komargodski:2011xv}. Their nonperturbative method has been further developed in 
\cite{Dymarsky:2013pqa,Farnsworth:2013osa,Dymarsky:2014zja} (however see \cite{Bzowski:2014qja}), 
and also similar argument was used to show perturbative equivalence in three dimensions \cite{Yonekura:2012kb}.

The aim of this paper is to complete the line of argument of \cite{Luty:2012ww}.
The result of \cite{Luty:2012ww} (which will be further developed in this paper) is that 
in four dimensional unitary scale invariant field theories, the Fourier transform of the trace $T=T^\mu_\mu$ of the energy-momentum tensor, $\tilde{T}(p)$,
vanishes when $p^2=0$. Then, we will show that the trace is given as $T=\partial^2 O$ where $O$ is a local operator.
By the improvement $T^{\mu\nu} \to T^{\mu\nu}-\frac{1}{3} (g^{\mu\nu}\partial^2-\partial^\mu\partial^\nu)O$, it becomes traceless
and hence the theory is conformal.

\paragraph{Dilaton amplitudes.}---
Let us briefly recall the argument of \cite{Luty:2012ww}.
We couple the theory to the metric of the form $g_{\mu\nu}=(1+\varphi)^2 \eta_{\mu\nu}$. We call the scalar $\varphi$ as dilaton.
As in \cite{Dymarsky:2013pqa} (see also \cite{Grinstein:2008qk}), $T_{\mu\nu}$ can be taken such that it is an eigenstate of the dilatation operator $D$,
up to possible mixing with a (generalized~\cite{Dymarsky:2013pqa}) dimension two operator $Y$ as
 $i[D,T]=x^\rho \partial_\rho  T+4 T+\partial^2 Y. $
 We couple the metric to this $T_{\mu\nu}$. The term containing $Y$ does not contribute to the dilaton amplitudes because
 we will impose the on-shell condition $p^2=0$ and it can be neglected. 
 
 We introduce a large kinetic term of the dilaton $\varphi$ by $\int (f^2/6) \sqrt{-g} R=\int f^2 (\partial \varphi)^2$, where $f$ is taken to be much larger
 than any other scale. Then we consider the scattering $\varphi(p_1) \varphi(p_2) \to \varphi(p_3) \varphi(p_4) $ of the massless dilations.
 We only consider the forward amplitudes $A(s) $ by taking $p_1=p_3$ and $p_2=p_4$, where $s=(p_1+p_2)^2$.
 By the scale invariance, $A(s)$ in the large $f$ limit is given by $A(s) = a s^2 /f^{4}$, where $a$ is a constant, since there is no counterterm for this amplitude
(see \cite{Luty:2012ww,Dymarsky:2013pqa}).

This amplitude has no branch cut on the complex $s$ plane, and hence by the optical theorem, we conclude that
all the amplitudes ($\varphi\varphi \to {\rm anything}$) vanish. 
The couplings of the dilaton to the theory is given by $\int \varphi T+\CO(\varphi^2)$. Therefore, in terms of operators, we get
\beq
\bra{\Phi} \Time\{\tilde{T}(p_1) \tilde{T}(p_2)\} +K(p_1,p_2) \tilde{T}'(p_1+p_2)) \ket{0}=0,\label{eq;dilatoncut}
\eeq
where $\bra{\Phi}$ is an arbitrary state, $\ket{0}$ is the vacuum state, 
$\Time$ is the time ordering, and $K(p_1,p_2) \tilde{T}'(p_1+p_2)$ is the contribution coming from 
the $\CO(\varphi^2)$ couplings $\varphi(p_1)\varphi(p_2) K(p_1,p_2) \tilde{T}'(p_1+p_2) $ in the Lagrangian, where $K(p_1,p_2) $ is a polynomial of $p_1 $ and $p_2$,
and $T'$ is some operator. 
$\Time\{\tilde{T}(p_1) \tilde{T}(p_2)\}$ is
the Fourier transform of $\Time\{{T}(x_1) {T}(x_2)\}$.

Now we are going to argue that $\bra{\Phi_1} \tilde{T}(p) \ket{\Phi_2}=0 $ for arbitrary states $\ket{\Phi_1}$ and $\ket{\Phi_2}$, as long as $p^2=0$.
In the following, it is implicit that we only consider connected diagrams, which is enough for our purpose.

For the moment, suppose that the theory has the S-matrix, at least when IR divergences are regularized. 
IR regularization might be physically realized by relevant deformation with very small mass parameters and/or
giving vevs to some operators of the theory. It is very likely that most theories have such IR regularization.

If the existence of the S-matrix is assumed,
the above result can be written as 
$\vev{\Phi: {\rm out} | \varphi\varphi :{\rm in}}=0$ at least when the energy scale is much larger than the IR cutoff, 
where $\Phi \neq 0$ represents a set of particles,
and $\bra{* : {\rm out}}$ and $\ket{* : {\rm in}}$ represent in and out states, respectively.
Taking $\Phi=\Phi'+\Phi''$ and using crossing symmetry, we get $\vev{\Phi',\varphi: {\rm out} |\Phi'', \varphi :{\rm in}}=0$.
Then by considering forward amplitudes  $\vev{\Phi', \varphi: {\rm out} |\Phi', \varphi :{\rm in}}=0$,
and again using the optical theorem, we get 
$\vev{\Psi: {\rm out} |\Phi', \varphi :{\rm in}}
=\bra{\Psi: {\rm out}} \tilde{T}(p)\ket{\Phi':{\rm in}}=0$
for arbitrary $\Psi \neq 0$. The cases $\Phi'=0$ and/or $\Psi=0$ can be treated separately and the same result holds.
Since each in and out state spans the Hilbert space,
we conclude that $\bra{\Phi_1} \tilde{T}(p) \ket{\Phi_2}=0$ holds.

In the above argument, the existence of the S-matrix was used only in the claim that
if Eq.~(\ref{eq;dilatoncut}) is satisfied for arbitrary $ \ket{\Phi}$, 
then for arbitrary  $\ket{\Phi'}$ and $\ket{\Phi''}$,
\beq
\bra{\Phi'} \Time\{\tilde{T}(p_1) \tilde{T}(p_2)\} +K(p_1,p_2) \tilde{T}'(p_1+p_2) \ket{\Phi''}=0, \label{eq:crossing}
\eeq
for $p_{1,2}^2=0$, $p_{1}^0>0$ and $p_2^0<0$.
We take this weak version of ``crossing symmetry'' as one of the assumptions of this paper, regardless of the existence of the S-matrix.

We have also used the optical theorem in the above argument, but that may be done without using the S-matrix.
In Eq.~\eqref{eq:crossing}, suppose that $\bra{\Phi'}$ has momentum $k$.
Define $P=p_1+k$. On the complex $P^0$ plane, there is a branch cut across real $P^0$ such that $P^0>|\vec{P}|$.
By taking discontinuity across the cut, 
we obtain
$
0=\bra{\Phi'} \tilde{T}(p_1) \tilde{T}(p_2) \ket{\Phi''}.
$
See the later discussion.
Note that the term $K{\tilde{T}}'$ does not contribute to the discontinuity and it drops out.
Then, we must have $\tilde{T}(p) \ket{\Phi}=0$ for $p^2=0$ in unitary theory.

\paragraph{Definition of operator.}---
We are going to argue that
\beq
O(x) \equiv  \frac{1}{\partial^{2}} T(x) ~~{\rm or~equivalently}~~\tilde{O}(p) \equiv \frac{-1}{p^2} \tilde{T}(p) \nonumber
\eeq
is a well-defined local operator. Our assumption is that the operator $O$ is a local operator if it satisfies the Wightman axioms \cite{Streater:1989vi}
(but we give up mathematical rigor).
The following needs to be checked:
(1) $O$ is an operator-valued distribution, or equivalently, Wightman correlation functions 
$\bra{0}O_1(x_1) \cdots  O(x)  \cdots O_n(x_n) \ket{0} $
are distributions.
(2) $O$ must satisfy $[O(x), O'(y)]=0$ with an arbitrary local operator $O'$ if $x-y$ is space-like.
Our aim is to verify these properties. Other necessary properties, such as Lorentz covariance, can be checked easily.

\paragraph{Prelimilary.}---
We will use two types of correlation functions: Wightman correlation function $\CW$ and
time ordered correlation function $\CT$.
They are defined as
\beq
&\CW(x_1,\cdots,x_n)=W(y_1,\cdots,y_{n-1}) \nonumber \\
&=\bra{0} O_1(x_1) \cdots O_n(x_n) \ket{0} , \nonumber \\
& \CT(x_1,\cdots,x_n)  
=\sum_\pi  \CW_\pi  \prod_{i=1}^{n-1}\theta(x^0_{\pi(i)}-x^0_{\pi(i+1)}), \nonumber 
\eeq
where $O_i~(i=1,\cdots,n)$ are arbitrary operators, $y_i=x_i - x_{i+1}$, 
$\theta$ is the step function, $\CW_\pi $ means the Wightman correlation function with operators $O_i(x_i)$ permuted by $\pi$ and the sum 
is over all permutations.
In the definition of $W(y_1,\cdots,y_{n-1})$ we have used translation invariance.

The Fourier transform of some quantity $A(x)$ is denoted as $\tilde{A}(p)= \int dx e^{ip x} A(x)$.
In particular, we define $\tilde{W}(q_i)= \int \prod_i d^4y_i e^{i q_i y_i} W(y_i) $, and the Fourier transforms of $\CW$ and  $W$ are related as 
\beq
\tilde{\CW}(p_i) &= (2\pi)^4 \delta^4(p_1+\cdots +p_n) \tilde{W} (q_1,\cdots,q_{n-1}), \nonumber \\
q_i &=p_1+\cdots+p_i=-(p_{i+1}+\cdots+p_n).
\eeq
Since the state $\tilde{O}(p_{j}) \cdots \tilde{O}(p_{n}) \ket{0}$ has momentum $P^\mu=-(p_{j}+\cdots+p_{n})^\mu=q^\mu_{j-1}$,
$\tilde{W}$ has support on the light cone $q_i^0 \geq |\vec{q}_i|$ (spectral condition).

From the above definition of $\CT$ and $\CW$, the relation between them in momentum space is given by
\beq
&\tilde{\CT} /(2\pi)^4 \delta^4(p_1+\cdots +p_n) \nonumber \\
&=  \prod_{i=1}^{n-1} \int_{|\vec{q}_i|}^\infty  \frac{d E_i}{2 \pi} \frac{i}{q^0_i -E_i +i \epsilon} \cdot  \tilde{W} ((E_i,\vec{q}_i))+\cdots,\label{eq:TWrelation}
\eeq
where $\epsilon>0$ is the usual Feynman $\epsilon$. We have only shown the contribution from the identity permutation $\pi(i)=i$,
and ellipses denote terms coming from other permutations.
Time-ordered correlation functions have
branch cuts located on the submanifolds of complex momentum space  
where $q_i^0$ is real and $q_i^0 \geq |\vec{q}_i|$, 
and other submanifolds obtained from permutations of $p_i$.
From the above expression, we can see that Wightman correlation functions are obtained from time-ordered correlation functions
by taking discontinuity across  the cuts, by using the formula $(q^0_i +i \epsilon-E_i )^{-1}-(q^0_i-i \epsilon -E_i )^{-1}=-2\pi i \delta(q_i^0 -E)$.

\begin{figure*}[] 
\includegraphics[scale=0.26]{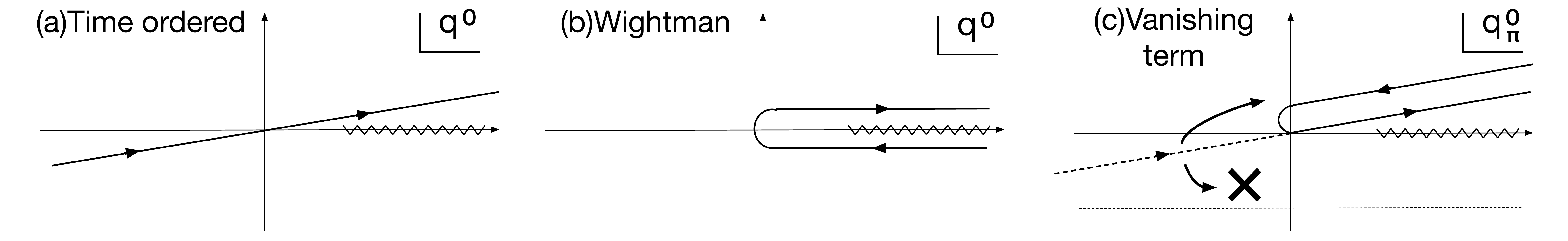}
\caption{\label{fig:cont} 
Contours in the  complex $q^0$ plane to obtain (a)time-ordered or (b)Wightman correlation functions.
Branch cuts are shown by zigzag lines. (c)Some of the branch cuts corresponding to a ``wrong'' permutation $\pi$ cannot be wrapped.
}
\end{figure*}

More precise procedure of obtaining $\CW$ from $\CT$ is the following.
Let us consider the case that the time variables are ordered as $x_1^0>x_2^0>\cdots>x_n^0$.
Then $\CW(x_i)$ and $\CT(x_i)$ are the same by definition. 
The coordinate space $\CT(x_i)$ is obtained from the momentum space $\tilde{\CT}(p_i)$ by the integration
$\CT(x_i)=\int \prod_i (d^4 p_i/(2 \pi)^4) e^{-i p_i x_i} \tilde{\CT}(p_i)$ where we take the integration contour as in Fig.\ref{fig:cont}-(a).
Now, we try to deform the integrations over $q^0_i~(i=1,\cdots,n-1)$ so that the integration 
contours shrink to zero as far as possible, 
other than possibly wrapping branch cuts.
For the moment, let us neglect branch cuts coming from ellipses in Eq.~\eqref{eq:TWrelation}.
Then we can deform the contours as in Fig.\ref{fig:cont}-(b). Note that because we are taking $y_i=x_i-x_{i+1} >0$,
the exponent of $e^{-i q_i y_i}$ can be taken to satisfy $\Re (-i q^0_i y^0_i)  <\epsilon \to 0$ in the process of contour deformation from 
Fig.\ref{fig:cont}-(a) to Fig.\ref{fig:cont}-(b). 
Thus the integral is convergent 
and this analytic continuation is justified.
In Fig.\ref{fig:cont}-(b), the integral can be evaluated by taking discontinuity across the real positive $q_i^0$. 
This discontinuity of $\tilde{\CT}(p_i) $ is precisely the momentum space Wightman function $\tilde{\CW} (p_i)$
discussed in the previous paragraph. Note that we are integrating over real $n-1$ dimensional surface if we fix $\vec{q}_i$,
and this surface is folded $n-1$ times. The integral is that of a holomorphic top form, so the $n-1$ dimensional surface 
can be deformed freely.

In the process of analytic continuation, we also encounter other branch cuts associated to the terms denoted by ellipses in Eq.~\eqref{eq:TWrelation}.
Let us define $y_{\pi, i}=x_{\pi(i)}-x_{\pi(i+1)}$ and $q_{\pi, i}=p_{\pi(1)}+\cdots+p_{\pi(i)}$ corresponding to a permutation $\pi$.
Then the branch cuts are located on $q_{\pi, i}^0  \geq |\vec{q}_{\pi, i}|$. 
Let $J$ and $K$ be sets such that for $j \in J$ and $k \in K$, $y_{\pi, j}>0$ and $y_{\pi, k}<0$.
Because of the ordering $x_1^0>x_2^0>\cdots>x_n^0$, $K$ is not an empty set if $\pi$ is not the identity element.
We have taken the contour deformation such that 
$\Re (-i \sum_i p_i x_i)=\Re (-i \sum_i q_{\pi,i} y_{\pi,i} )<0$, so it is impossible to
go to a point at which $\Im q_{\pi, j}>0~(j \in J)$ and $\Im q_{\pi, k} <0~(k \in K)$.
Then we can never reach the configuration in which $ q_{\pi, j} \to \Re q_{\pi, j} +i0 $ and $q_{\pi, k} \to \Re q_{\pi, k} -i0 $
with $\Re q_{\pi, j}$ and $\Re q_{\pi, k}$ extending to positive infinity. 
(The only exception is
$q_{\pi, i} \to \Re q_{\pi, i} +i0$ for all $i=1,\cdots,n-1$
even if $K=\emptyset $, because it is the initial starting point.)   
Thus it is impossible pick up the discontinuity of all the variables $q_{\pi, i}~(i=1,\cdots,n-1)$.
Recalling that we are folding the integration surface $n-1$ times,
we conclude that the term associated to $\pi$ vanishes (see Fig.~\ref{fig:cont}-(c)).
This must be so because $\CW(x_i)=\CT(x_i)$ for $x_1^0>x_2^0>\cdots>x_n^0$ by definition.

Note that the restriction $y_i>0$ was necessary only in the process of the analytic continuation
from Fig.\ref{fig:cont}-(a) to Fig.\ref{fig:cont}-(b), i.e., in obtaining $\CW(x_i)$ from $\CT(x_i)$,
and is not necessary in each integral Fig.\ref{fig:cont}-(a) or Fig.\ref{fig:cont}-(b) alone.

\paragraph{Dilaton result.}---
Let us interpret the result obtained before.
Consider a Wightman correlation function
\beq
&\tilde{\CW}_T 
=\bra{0}\tilde{O}_1(p_1) \cdots  \tilde{T}(p)  \cdots \tilde{O}_n(p_n) \ket{0}.
\eeq
We focus on the connected component.
Other than the momentum conservation delta function $(2\pi)^4 \delta^4(p_1+\cdots +p_n)$,
it is given by a function \footnote{For simplicity, 
we assume that integration of these functions with test functions are absolutely convergent near singular loci.
If the theory has genuine one particle states, they can also contain 
distributions $\delta(q^2)$ and $P(1/q^2)$,
but we believe that does not affect our discussion. More strong singularities may be excluded by unitarity.} 
of the momenta which is obtained by taking discontinuity of the corresponding time-ordered correlation function. 
From the result of the dilaton argument, this function must vanish when $p^2=0$.

If $\tilde{\CW}_T$ vanishes on $p^2=0$, the corresponding time-ordered correlation function $\tilde{\CT}_T $
can be defined to vanish, i.e., $\tilde{\CT}_{T}|_{p^2=0}=0$. 
Suppose $\tilde{\CT}_T $ does not vanish on $p^2=0$ and denote the non-vanishing term as $\tilde{\CT}'_T $.
That term does not contribute to $\tilde{\CW}_T$ when we obtain $\tilde{\CW}_T$ from $\tilde{\CT}_T$.
In the coordinate space, if time variables are ordered as $x^0_{\pi(1)}>\cdots>x>\cdots>x^0_{\pi(n)}$, then 
${\CT}_T$ coincides with ${\CW}_T$  with operators permuted correspondingly. This means that ${\CT}'_T $
is nonzero only if $x^0=x^0_i$ or $x^0_i = x^0_j$ for some $i$ and $j$. By Lorentz invariance, it has support on $x^\mu=x^\mu_i$ or $x^\mu_i = x^\mu_j$.
Such contact terms can be subtracted by counterterms for time-ordered correlation functions \footnote{Strictly speaking, 
``time-ordered correlation functions'' may be regarded as analytic continuation
of Euclidean correlation functions. There, values at $x_i = x_j$ are not defined (even as distribution) in 
the Osterwalder-Schrader axioms,
and these values are unimportant in reproducing Wightman distributions \cite{Osterwalder:1973dx,Osterwalder:1974tc}.
However, one should not misunderstand that the contact terms appearing in
Ward identities or anomalies are meaningless.
There, we take derivatives (or trace in the case of trace anomaly)
of correlation functions and get contributions which have support on $x_i = x_j$. 
These contact terms contain information of nonlocal behavior $x_i \neq x_j$ of the original correlation functions before taking derivatives or trace,
and therefore meaningful. 
For example, $ g^{\mu\nu}(p_\mu p_\nu/p^2)=1$ is local, but $(p_\mu p_\nu/p^2)$ is not.
Our interest in this paper is not the original operator $T_{\mu\nu}$, but $T$ itself, so it is our freedom to drop contact terms.
If we restore those contact terms, our result is written as $g^{\mu\nu}T_{\mu\nu}=\partial^2 O+({\rm contact~terms})$ 
in time-ordered 
correlation 
functions, which is the usual statement of Ward identities for the trace of $T_{\mu\nu}$.
We stress that there is no notion of contact terms in the Wightman's case.
}.

\paragraph{Well-definedness.}---
Before going to discuss $\tilde{O}(p)=-p^{-2} \tilde{T}(p)$,
we remark that a slightly deformed operator
\beq
\tilde{O}_\eta (p)=\frac{-1}{p^{2} +i \eta} \tilde{T}(p),~~~(\eta \in {\mathbb C}) \label{eq:deformedop}
\eeq
is a perfectly well-defined (though neither local nor hermitian) operator-valued distribution if $\Re \eta \neq 0$,
because for an arbitrary test function $\tilde{f}(p)$, the function $\tilde{g}(p)=\tilde{f}(p)/(p^{2} +i \eta)$
is also a test function (i.e., infinitely differentiable and rapidly decreasing function
$\sup_{p \in {\mathbb R}^{1,3}} (|p|^\alpha |(\partial)^{\beta} \tilde{g}|) < +\infty$ for arbitrary $\alpha, \beta \geq 0$; see \cite{Streater:1989vi}).
The question is whether we can take $\eta \to 0$ or not.

Let us roughly estimate the behavior of $\tilde{\CT}_T$ and $\tilde{\CW}_T$ for small $p^2$. If it were an analytic function of momenta, it would have 
a Taylor expansion 
and we would get $\tilde{\CT}_T = \CO(p^2)$.
Actually $\tilde{\CT}_T$ is not quite an analytic function, but the non-analyticity about $p^2$ is assumed to come from on-shell processes
with the discontinuity proportional to $\sum_\Phi \bra{0} \Time \{\tilde{O}_1(p_1) \cdots \tilde{O}_1(p_n)\} \ket{\Phi}\bra{\Phi} \tilde{T}(p)  \ket{0}$.
(This may be shown in the same way as the derivation of the Lehmann-Symanzik-Zimmermann reduction formula.
See e.g. section 10.2 of \cite{Weinberg:1995mt}.)
Let $\tilde{S}(p)=\int \prod_i d^4p_i \delta^4(p-\sum_i p_i) f(p_1,\cdots,p_n) \Time \{\tilde{O}_1(p_1) \cdots \tilde{O}_1(p_n)\}$ be a smeared
product of operators with total momentum $p$. Then, by unitarity we have
$|\rho_{ST}| \leq \sqrt{\rho_{SS} \rho_{TT} } $, where
\beq
\rho_{AB}(p) \equiv \left(\sum_\Phi \bra{0} \tilde{A}(p) \ket{\Phi} \bra{\Phi} \tilde{B}^\dagger(p')   \ket{0} \right)/ (2\pi)^4\delta^4(p+p'). \nonumber
\eeq
The $\rho_{SS}$ is a distribution about $p$ satisfying $\rho_{SS} \geq 0$ by unitarity. 
Then the singularity must be weaker than $p^{-2}$,  $\rho_{SS} < \CO(|p^{-2}|)$ for small $p^2$.
On the other hand, $\rho_{TT} = \CO(p^{2(\Delta_T-2)})$ where $\Delta_T=4$ is the scaling dimension of $T$.
Thus we estimate $|\tilde{\CW}_T| \sim |\tilde{\CT}_T | \sim |\rho_{ST}|  < \CO(|p^2|^{(\Delta_T-3)/2})$ from the non-analytic contribution, 
which dominates over analytic contributions $\CO(p^2)$.

Now consider the correlation function
\beq
&\tilde{\CW}_O
=\bra{0}\tilde{O}_1(p_1) \cdots  \tilde{O}(p)  \cdots \tilde{O}_n(p_n) \ket{0}=\frac{-1}{p^2} \tilde{\CW}_T.
\eeq
By the above estimate, we get $|\tilde{\CW}_O| < \CO(|p^2|^{(\Delta_T-5)/2})$.
Then the integral $\int d^4 p \tilde{f} (p) \tilde{\CW}_O(p)$ converges for an arbitrary test function $\tilde{f} (p)$ since the singularity is weak.
We conclude that $\tilde{\CW}_O$ is a well-defined distribution. 

\paragraph{Locality.}---
Our final task is to show the locality of the operator $O$.
It is instructive to consider the more general operator $O_\eta$ introduced in Eq.~(\ref{eq:deformedop}).
Define the time-ordered correlation function 
\beq
\tilde{\CT}_{O_\eta}=\bra{0} \Time \{\tilde{O}_1(p_1) \cdots  \tilde{O}_\eta(p)  \cdots \tilde{O}_n(p_n) \} \ket{0}.
\eeq
One might think it is just $-(p^2+i\eta)^{-1}\tilde{\CT}_{T}$, but it is not the case.
Our definition is that $\tilde{\CW}_{O_\eta}=-(p^2+i\eta)^{-1}\tilde{\CW}_{T}$, and 
the coordinate space Wightman function ${\CW}_{O_\eta}$ is given by the integral of $-(p^2+i\eta)^{-1}\tilde{\CT}_{T} e^{-ipx-i\sum p_i x_i}$
along the contour shown in Fig.\ref{fig:cont}-(b).
However, to get time-ordered correlation functions,
we have to do analytic continuation from Fig.\ref{fig:cont}-(b) back to Fig.\ref{fig:cont}-(a).
Then, $(p^2+i\eta)^{-1}$ introduces a new pole and we have to pick up the residue which is proportional to $\tilde{\CT}_{T}|_{p^2=-i\eta}$.
This contribution from the pole may depend on how the time variables are ordered.
However, this residue becomes zero when $\eta \to 0$. Therefore, 
we simply get $\tilde{\CT}_{O}=-p^{-2}\tilde{\CT}_{T}$ in this limit.
In particular, it is Lorentz invariant.

The Lorentz invariance of ${\CT}_{O}$ is highly nontrivial. 
If $x-x'$ is space-like, $\Time \{O(x)O'(x')\}=O(x)O'(x')$ in one Lorentz frame and $\Time \{O(x)O'(x')\}=O'(x') O(x)$ in another one.
Therefore, we must have $O(x)O'(x')=O'(x')O(x)$, which establishes the locality condition $[O(x),O'(x')]=0$.
(Strictly speaking, we use the fact that the locality condition of the Wightman axioms is equivalent to
the permutation invariance of Euclidean functions of 
the Osterwalder-Schrader axioms~\cite{Osterwalder:1973dx,Osterwalder:1974tc}.)


\paragraph{Comment on compact scalar.}---
In $d$-dimensional space-time with $d>2$, a free $d-2$ form gauge field or equivalently its dual compact scalar is often discussed as a counterexample 
to the equivalence of scale and conformal invariance. 
However, we must be careful about 
what we mean by ``quantum field theory'' to see whether this example is really a counterexample or not.
By taking the Wightman axioms as the definition of quantum field theory, we now argue that this is not a counterexample.

A compact scalar $\phi$ has the ${\mathbb Z}$ gauge invariance 
$\phi \cong \phi+f n$, where $n \in {\mathbb Z}$ and $f$ is a constant. 
However, in ${\mathbb R}^{1,3}$, cluster decomposition \cite{Streater:1989vi} tells us that we have to choose a specific vacuum
expectation value of $\phi$, say $\vev{\phi}=0$ mod $f {\mathbb Z}$. 
See section~19 of \cite{Weinberg:1996kr} for detailed discussions.
Then we can just fix the  spontaneously broken ${\mathbb Z}$
gauge symmetry by ``unitarity gauge'' $\vev{\phi}=0$. After choosing this, quantization of the field $\phi$ is straightforward;
there is no way to quantize it in flat Minkowski space ${\mathbb R}^{1,3}$ other than doing just the textbook quantization of the usual massless scalar.
In particular, the Hilbert space is just a Fock space of the free scalar. 
(See \cite{Coleman:1973ci} for the explanation of how this argument fails in two dimensions.)
We might not initially include $\phi^m~(m=1,2,\cdots)$ in the set of 
local operators of the theory. But
after fixing the unitarity gauge $\vev{\phi}=0$, there is no reason to exclude them.
We conclude that the compact scalar is completely the same as the usual non-compact scalar as long as we are concerned with 
the Hilbert space and local operators in ${\mathbb R}^{1,3}$, which are the ingredients of the Wightman axioms.
This is consistent with the result of this paper.
The conclusion is unchanged even if we take the limit $f \to 0$, i.e., ${\mathbb R}$ gauge symmetry $\phi \to \phi+c,~c \in {\mathbb R}$.

Now we would like to consider two things which are not taken into account in the Wightman axioms, 
and hence out of the scope of this paper.
One is to replace ${\mathbb R}^{1,3}$ with a manifold $M$ of nontrivial topology.
We have a winding mode
which is absent in the non-compact scalar. However, note that the scale invariance is explicitly broken by the scale $f$ which enters into the physical spectrum in this case. 
The winding modes are also important in entanglement entropy \cite{Agon:2013iva} which is related to a partition function on nontrivial space
by replica trick.

The other thing is to include nonlocal operators such as line or surface operators.
(A similar problem was discussed in \cite{Aharony:2013hda} where two theories are considered which 
are different only by nonlocal operators.)
For example, by using two-form gauge field $B$ in four dimensions, we can have a surface operator $\exp (i n \int_S B )$ where
$S$ is some two dimensional submanifold and $n \in {\mathbb Z}$. Then, going around the surface, the dual compact scalar is not single valued, but behaves as 
$\phi \to \phi+f n$. In the presence of this surface operator, the space-time might appropriately be considered not as ${\mathbb R}^{1,3}$
 but as ${\mathbb R}^{1,3} \setminus S $.
Also in this case, the scale invariance is not preserved by this operator, since the behavior $\phi \to \phi+f n$
around the surface cannot be preserved under scaling transformation for $n \in {\mathbb Z}$.

By the above reasons, we do not consider the compact scalar as a counterexample to the equivalence of scale and conformal invariance.
Depending on our definition of quantum field theory, both scale and conformal invariance are preserved or both of them are broken.
However, the theory with ${\mathbb R}$ gauge symmetry with nonlocal operators or nontrivial space-time $M$ 
could be a counterexample~\cite{Dymarsky:2013pqa}, 
which is analogous to theories with continuous operator spectrum in two-dimensions~\cite{Polchinski:1987dy}, since it has
continuous nonlocal operators or continuous spectrum in $M$.

\paragraph{Acknowledgement.}---
This work is somewhat influenced by K.~Usui and it is a pleasure to thank him.
The author would also like to thank T.~Nishioka, Y.~Oshima, and N.~Seiberg for helpful discussions.
The work of K.Y. is supported in part by NSF Grant PHY-0969448.


\bibliography{ref}

\end{document}